\documentclass[conference]{IEEEtran}
\IEEEoverridecommandlockouts 
\usepackage{cite}
\usepackage{amsmath,amssymb,amsfonts}
\usepackage{algorithmic}
\usepackage{graphicx}
\usepackage{textcomp}
\usepackage{xcolor}

\def\BibTeX{{\rm B\kern-.05em{\sc i\kern-.025em b}\kern-.08em
    T\kern-.1667em\lower.7ex\hbox{E}\kern-.125emX}}
\begin{document}

\title{AirDnD - Asynchronous In-Range Dynamic and Distributed Network Orchestration Framework\\
\thanks{This work has been supported by Swedish Foundation for Strategic Research (SSF), Grant Number FUS21-0004 SAICOM.}
}

\author{\IEEEauthorblockN{Malsha Ashani Mahawatta Dona}
\IEEEauthorblockA{
\textit{University of Gothenburg, Sweden}\\
malsha.mahawatta@gu.se}

\and
\IEEEauthorblockN{Christian Berger}
\IEEEauthorblockA{
\textit{University of Gothenburg, Sweden}\\
christian.berger@gu.se}
\and

\IEEEauthorblockN{Yinan Yu}
\IEEEauthorblockA{
\textit{Chalmers University of Technology, Sweden}\\
yinan@chalmers.se}
}
\maketitle
\begin{abstract}
The increasing usage of IoT devices has generated an extensive volume of data which resulted in the establishment of data centers with well-structured computing infrastructure. Reducing underutilized resources of such data centers can be achieved by monitoring the tasks and offloading them across various compute units. This approach can also be used in mini mobile data ponds generated by edge devices and smart vehicles. This research aims to improve and utilize the usage of computing resources in distributed edge devices by forming a dynamic mesh network. The nodes in the mesh network shall share their computing tasks with another node that possesses unused computing resources. This proposed method ensures the minimization of data transfer between entities. The proposed AirDnD vision will be applied to a practical scenario relevant to an autonomous vehicle that approaches an intersection commonly known as ``looking around the corner'' in related literature, collecting essential computational results from nearby vehicles to enhance its perception. The proposed solution consists of three models that transform growing amounts of geographically distributed edge devices into a living organism. 
\end{abstract}

\section{Introduction} 

AirDnD is a vision that brings the idea behind Airbnb\textsuperscript{\textregistered} into the domain of distributed systems. Airbnb is a significant hotel chain around the world without owning physical properties, yet bringing people who seek accommodation together with people who wish to rent their unused properties\cite{6859816}. A similar concept is envisioned for the growing amount of distributed embedded systems that spread across different geographical areas to achieve better utilization of the excess computing resources. 

The increasing usage of IoT devices has resulted in generating an extensive volume of data which initiated the existence of data centers with well-structured computing infrastructure. The data center operators strive to reduce underutilized resources by monitoring the tasks and disseminating them between different compute units to improve performance while utilizing the excess compute resources. The concepts of Serverless Computing and Function-as-a-Service (FaaS) \cite{Faas} are widely used in such contexts. Smart orchestration of computing resources with the support of AI/ML models helps \cite{AIML} to deploy such concepts by considering the physical properties of data centers, such as computing nodes, communication infrastructure, and various properties of the task. Similarly, edge devices and smart vehicles with compute resources are also becoming mini data ponds due to the extensive volumes of data they collect.

\subsection{Motivation}

The proposed research agenda is to improve and utilize the usage of computing resources in distributed edge devices, including modern vehicle fleets. The main motivation behind AirDnD is to enable better utilization of resources in computing devices that are geographically distributed by following the concepts used in immobile data centers. The geographically distributed, wireless embedded systems can be considered mobile data centers, which constantly reshape and grow with the data collected through different sensory interfaces. Developing and evaluating models for different layers of the proposed vision to capture, analyze, and predict resource allocation is required to enable better utilization of excess computing resources

Even though the advancements in wireless communication, including 5G and later versions, along with the increased computing capabilities of IoT devices, have enabled the frequent use of AI/ML engineering processes which are largely dependent on centralized datasets. However, mobile data ponds grow daily, making them immobile and stationary. Therefore, the 5G and beyond networks which provide growing bandwidths and reduced latencies, must be better used than in transferring millions of data back and forth between the centralized servers and edge devices. This articulates the need of having the aforementioned model with the capability of capturing the properties of such data ponds and disseminating the tasks. Hence, we envision that edge devices must spontaneously form a dynamic mesh network for a certain time period to execute the smart orchestration avoiding cellular communication. 

The evaluation of AirDnD will be based on a practical scenario relevant to an autonomous vehicle that approaches an intersection. The approaching vehicle is in need of a wider view of the intersection to achieve the safety of road users. This application scenario is commonly known as ``looking around the corner'' \cite{5548138} in modern research areas.

By applying AirDnD in the said use case, the data will remain where they have been generated while the computing task that describes what is needed to be done will be exchanged to the data pond. The proposed solution consists of the below models in different layers. 
\\
\textbf{Model 1: Network Description} To describe the spontaneously forming and dissolving dynamic mesh network. 
\\
\textbf{Model 2: Task Description} To describe the task offloading. The task needs to be formal and abstract in a way that it could work on the receiving node.
\\
\textbf{Model 3: Data Description} To describe the type and the quality of data that shall be required by the exchanged compute task. 

\section{Research Plan}

Our vision is to develop and evaluate models that help to describe and reason about transforming geographically distributed edge devices into a living organism where computing and communication will live in symbiosis despite their heterogeneities.

\begin{figure}
\centerline{\includegraphics[width=\linewidth]{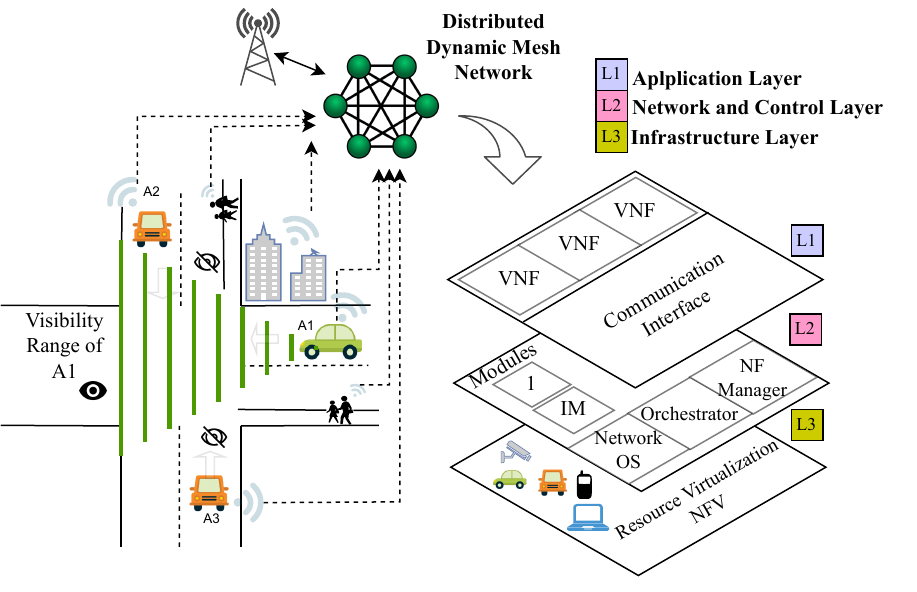}} %
\caption{The proposed vision of AirDnD}
\label{fig:architecture}
\end{figure}

\subsection{Research Methods}
The proposed vision will be divided into smaller research activities to establish a research plan.
\\
\textbf{Goal 1:} Explore the existing models and architecture in the existing literature to identify the gaps, limitations, and trends.
\\
\textbf{Goal 2:} Formulate the properties and the quality of the network and its dynamicity to systematically identify the functional and non-functional properties of the network. Focus areas are computing capabilities of the receivers, data quality, network parameters, and non-functional properties such as integrity, privacy, and trust.
\\
\textbf{Goal 3:} Models to describe what type and quality data is needed to successfully offload compute tasks on such mobile data centers (data expected for the computations and the results after the computation). Semantic protocols which enable communication between heterogeneous systems will also be established to allow time-critical collaboration between distributed systems. 

The following research questions were formulated with respect to the case study ``looking around the corner'' scenario. This case study will be used to evaluate the system in both simulated environments and in scaled vehicles from \cite{revere}. 
\begin{itemize}
    \item \textbf{RQ 1:} What qualities and properties must be considered when selecting the computing nodes?
    \item \textbf{RQ 2:} How to offload the tasks?
    \item \textbf{RQ 3:} How to handle the computation? (Feasibility, Privacy, integrity, and trust related to intellectual properties)
\end{itemize}
\subsection{Initial Results}
An analysis was conducted on the areas including Software Defined Networks (SDN) and  Network Function Virtualization (NFV) based on the SLR published by Bonfim et al.~\cite{10.1145/3172866} to identify the gaps and trends in the relevant literature. The areas such as smart orchestration and distributed NFV were identified as trending from the analysis. The results of this analysis played a vital role in understanding the necessity of conducting research in the fields such as smart orchestration with distributed NFV models. In addition to that, related literature \cite{Decloud,Geo,Auction} was studied to identify the limitations and gaps in similar concepts. The said literature mainly focuses on allocation and deallocation algorithms of computing resources, whereas the formation of a spontaneous dynamic mesh network is not addressed so far. The proposed AirDnD concept envisions an innovative method of utilizing the usage of computing resources in distributed edge devices by considering the identified gaps, limitations, and trends in the current research domain.

\subsection{Challenges}
\begin{itemize}
    \item The dynamicity of the network and computing tasks
    \item Modeling a scalable network
    \item Ensuring data quality, privacy, and security
    \item Delivering a trustworthy system enabling computing nodes to share their instructions and computing power
\end{itemize}


\bibliography{references}
\bibliographystyle{IEEEtran}
\vspace{12pt}

\end{document}